# Decoherence-free measurement of wavefunction collapse with interferometers in quantum superpositions

*Garrelt Quandt-Wiese*
*Schlesierstr. 16*
*64297 Darmstadt*
*Germany*
*garrelt@quandt-wiese.de*
*www.quandt-wiese.de*

A novel approach for measuring wavefunction collapse is proposed which, unlike interferometric measurements, is not affected by decoherence. A mirror of a Michelson interferometer is transferred into a quantum superposition, where the decay of the mirror superposition is measured by the fact that it affects the probability of detecting photons in the interferometer. The experiment can be realised with currently available technology for photon generation and detection and is suitable for testing collapse models, which is shown by calculating the expected outcomes for the gravity-based models of Diósi and Penrose.

## 1. Introduction

Direct interferometric measurements of wavefunction collapse face the problem of decoherence, which, like collapse, leads to the disappearance of interference patterns between the superposed states and thus obscures the effect to be measured. This problem becomes increasingly severe the more particles are involved. Although the number of particles in superpositions has already been significantly expanded—ranging from $2000$ atoms in the detection of superposed oligoporphyrin molecules [1], to $10^9$ electrons in superposed superconducting currents within SQUIDs [2], and even up to approximately $10^{16}$ atoms for acoustic waves in sapphire crystals coupled to a qubit [3]—the regime necessary to test established collapse models has yet to be reached.
In 2024, James Tagg et al. proposed a method for a decoherence-free measurement of wavefunction collapse using a Mach-Zehnder interferometer, where two of its mirrors are placed in quantum superposition, and their superpositions are measured indirectly through the intensity of the photon beam within the interferometer [4]. This approach could enable the measurement of superpositions of approximately $10^{21}$ atoms [4], providing a basis for testing collapse models, such as the gravity-based models of Diósi and Penrose [5], as well as the dynamical reduction models like Continuous Spontaneous Localization [6].

In this paper, we follow up Tagg's idea of a decoherence-free measurement of wavefunction collapse and analyse the experimental conditions necessary for its realization. For this, we consider a simplified version of his experiment, a Michelson interferometer in which only one mirror is placed in superposition. Our analysis shows (section 4) that the intensity measurement of the photon beam within the interferometer cannot be a classical one, but that the measurement's outcome must result from the reduction of a superposed state of the detector. This is given when single photons are detected, since for each photon detection a superposed state of the detector, consisting of a *'photon detected'* and a *'photon not detected'* state, reduces to one of these states. For this reason, we discuss single-photon detection for the analysis of the experiment and will show at the end of the paper that other measurement techniques could also be effective.

This work is structured as follows. In section 2 the experiment is introduced and in section 3 it is shown how it can be evaluated for testing collapse models. In section 4 the experimental conditions for measuring the collapse of the mirror superposition are discussed. In section 5, it is shown how the experiment can be implemented with current techniques for photon generation and detection. In section 6 the expected outcomes for the gravity-based models of Diósi and Penrose are calculated showing the feasibility of the approach for testing collapse models. In the appendix, it is shown how Tagg's experimental proposal can be evaluated with the results derived here.

## 2. Michelson interferometer in a quantum superposition

In this section, we introduce the experiment that serves as the basis for discussing the concept of a decoherence-free measurement of wavefunction collapse in this paper. Figure 1 shows the setup, in which the mirror in the upper arm of a Michelson interferometer is transferred into a two-state superposition of slightly different positions with help of a single-photon measurement. The two states of the mirror correspond to the cases in which the photon of this measurement is either detected or not detected by the single-photon avalanche photodiode (SPAD). In case of detection, the mirror is moved by the piezo actuator, which is driven by the avalanche current of the photodiode, while in the no-detection case the mirror is not moved. The intensities of the two states in which the mirror is moved, respectively not moved, $|c|^2_{mov}$ and $|c|^2_{\neg\, mov}$, depend on the transmission coefficient $T^2$ of the mirror in front of the photodiode, which couples out a portion of the photons, and on the detection probability $\eta$ (quantum efficiency) of the photodiode. For the intensities of the two states, we get:

$$\begin{aligned} |c|^2_{mov} &= T^2\eta \\ |c|^2_{\neg\, mov} &= 1 - T^2\eta \end{aligned} \quad . \tag{1}$$

This quantum superposition of the mirror decays after some time, either in favour of the state in which the mirror is moved by the piezo actuator with $\Delta s(t)$, or the state in which the mirror remains at rest ($\Delta s = 0$), whereby the reduction probabilities for these final states are given by $|c|^2_{mov}$ and $|c|^2_{\neg\, mov}$ according to Born's rule.

Before we examine in the next section how the quantum superposition of the mirror can be measured with this setup, we first consider the classical case in which the mirror is located at a unique position $\Delta s$. For the following derivation, we assume that the laser of the Michelson interferometer emits single photons with a photon emission rate $\dot{N}_{in}$ (see figure 1), which are detected by the detector of the interferometer. The

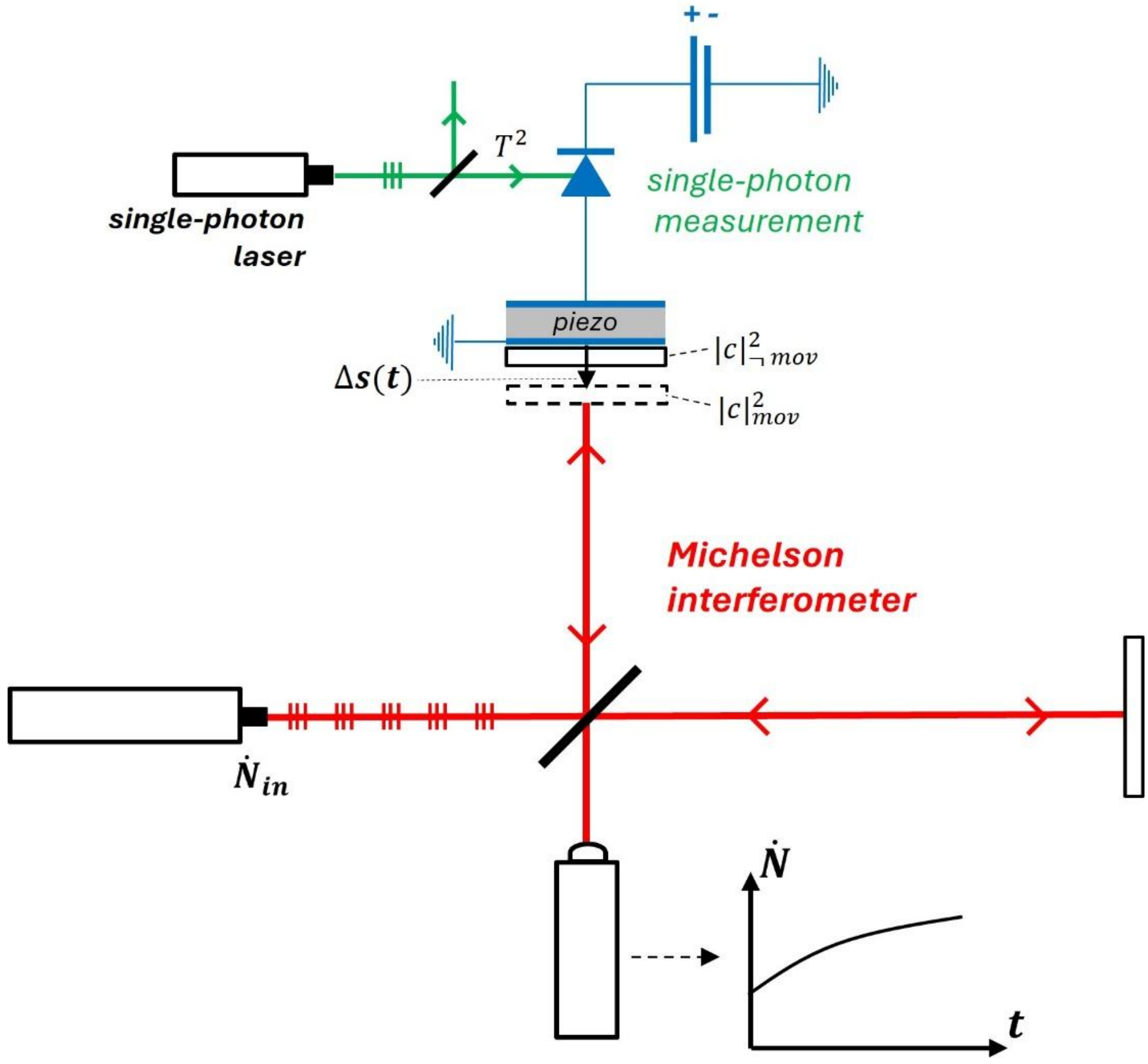

**Figure 1.** Setup for a decoherence-free measurement of wavefunction collapse (see text for details).

probability $p_{ph}$ to detect a photon is given as a function of the mirror displacement $\Delta s$ by:

$$p_{ph}(\Delta s) = \alpha \, sin^2\left(2\pi \tfrac{\Delta s}{\lambda} + \phi_0\right) + \beta \quad , \tag{2}$$

where $\lambda$ is the wavelength of the laser of the Michelson interferometer, and the constants $\alpha$, $\beta$ and $\phi_0$ depend on the arm lengths and the reflection coefficient of the semi-transparent mirror of the Michelson interferometer, and on the photon detection probability of the detector. For arm lengths differing by $\lambda/4$, lossless beam guidance in the Michelson interferometer and a photon detection probability of 100%, one has $\alpha = 1$, $\beta = 0$ and $\phi_0 = 45°$.

The detector of the Michelson interferometer records the photon detection rate $\dot{N}$ over time, as shown in figure 1, where the course $\dot{N}(t)$ depends on the mirror displacement $\Delta s(t)$ by:

$$\dot{N}(t) = \dot{N}_{in} \, p_{ph}(\Delta s(t)) \quad . \tag{3}$$

## 3. Evaluation of the experiment

In this section, we examine what results are expected when the mirror of the Michelson interferometer is in superposition and show how the experiment can be evaluated for testing collapse models by determining the decay rate of the superposed mirror.

When the mirror of the Michelson interferometer is in a quantum superposition, a photon detection in the Michelson interferometer leads to a four-state superposition, where the four states are given by the combinations *'mirror moved'*, *'mirror not moved'* and *'photon detected'* (by the detector of the Michelson interferometer!) and *'photon not detected'*. To simplify the discussion of the experiment, we assume that the lifetime of the photon detection $T_{ph-det}$ , i.e. the lifetime of the superposition of the *'photon detected'* and *'photon not detected'* states, is much smaller than the lifetime of the mirror superposition $T_{mi}$:

$$T_{ph-det} \ll T_{mi} \ , \tag{4}$$

and that $T_{ph-det}$ is also much smaller than the time separation $T_{in}$ between successive incoming photons in the Michelson interferometer ($T_{in} = 1/\dot{N}_{in}$; for $\dot{N}_{in}$ see figure 1):

$$T_{ph-det} \ll T_{in} \ , \tag{5}$$

which prevents successive photon detections from being in superposition at the same time. Under these assumptions, the probability of measuring a photon in the Michelson interferometer is given by:

$$p_{ph\,sup}(\Delta s) = |c|^2_{mov}\, p_{ph}(\Delta s) + |c|^2_{\neg\, mov}\, p_{ph}(0) \ \ , \tag{6}$$

and the photon detection rate is expected to be:

$$\dot{N}_{sup}(t) = \dot{N}_{in}\ (|c|^2_{mov}\, p_{ph}(\Delta s(t)) + |c|^2_{\neg\, mov}\, p_{ph}(0)\ ) \ \ . \tag{7}$$

The probability $P_{sup}$ that the mirror is still in superposition and the probabilities $P_{mov}$ and $P_{\neg\, mov}$ that the superposition has reduced to the state in which the mirror is moving respectively not moving, is governed by the decay rate of the mirror superposition $\Gamma(t)$ as follows:

$$\begin{aligned} \frac{dP_{sup}}{dt} &= -\Gamma(t)\ P_{sup}(t) \\ \frac{dP_{mov}}{dt} &= \ \ |c|^2_{mov}\Gamma(t)\ P_{sup}(t) \ \ . \\ \frac{dP_{\neg\, mov}}{dt} &= |c|^2_{\neg\, mov}\Gamma(t)\ P_{sup}(t) \end{aligned} \tag{8}$$

Depending on whether the mirror is still in superposition or has reduced to one of the two states, we expect the following photon detection rates:

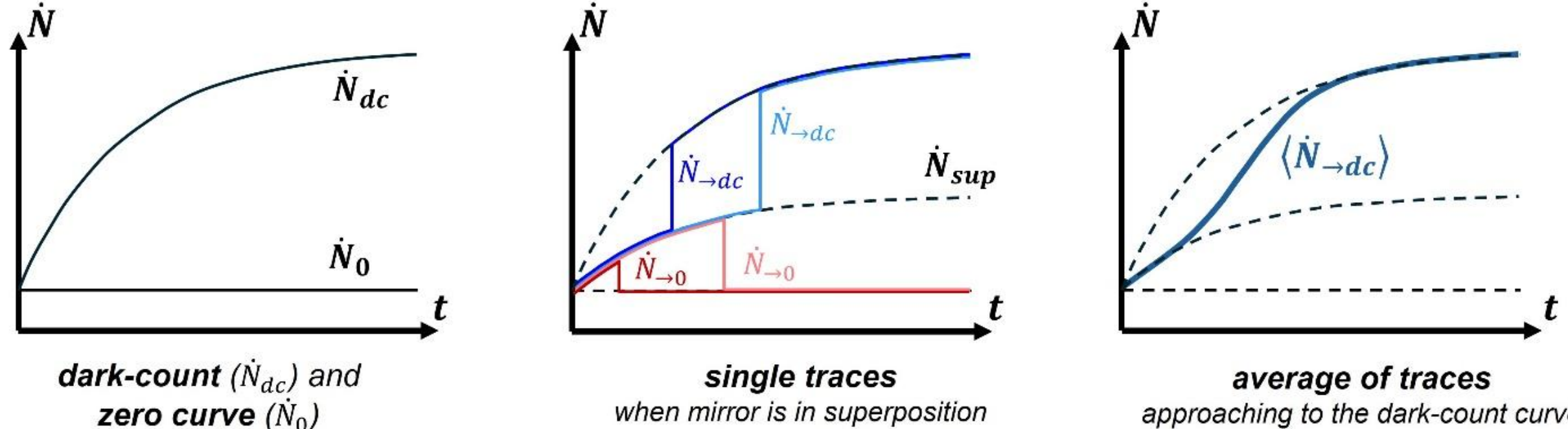

**Figure 2.** Photon detection rate $\dot{N}$ versus time measured by the detector of the Michelson interferometer in figure 1.

*Left:* Dark-count curve (when the photodiode spontaneously releases an avalanche current) and zero curve (when no photon is measured).

*Centre:* Individual traces, when the mirror is in a superposition. The traces initially follow the $\dot{N}_{sup}$-curve and then jump, due to collapse, at a random point in time to either the dark-count or the zero curve.

*Right:* Average of all traces that jump to the dark count curve after some time.

$$\dot{N}(t) = \begin{cases} \dot{N}_{sup}(t) & = \dot{N}_{in}\left(|c|^2_{mov}\, p_{ph}(\Delta s(t) + |c|^2_{\neg\, mov}\, p_{ph}(0)\right) & \textit{with } P_{sup}(t) \\ \dot{N}_{dc}(t) & \equiv \dot{N}_{in}\; p_{ph}(\Delta s(t)) & \textit{with } P_{mov}(t) \\ \dot{N}_0(t) & \equiv \dot{N}_{in}\; p_{ph}(0) & \textit{with } P_{\neg\, mov}(t) \end{cases} \quad . \tag{9}$$

The photon detection rate for the case that the superposition has been reduced to the state in which the mirror is moving is denoted here by $\dot{N}_{dc}(t)$, where *'dc'* stands for *'dark count'*, since this course is also expected when the photodiode spontaneously releases an avalanche current (without an incoming photon), which is denoted as a dark count event. The photon detection rate for the case in which the superposition has been reduced to the state in which the mirror is not moving is denoted as the *zero curve* $\dot{N}_0$, since the same rate is expected when the single-photon detector of our experiment does not emit a photon. The left part of figure 2 shows exemplarily the dark-count curve $\dot{N}_{dc}(t)$, and the zero curve $\dot{N}_0$. Due to the charging of the piezo actuator by the avalanche current of the photodiode, the displacement $\Delta s(t)$ and thus also $\dot{N}_{dc}(t)$ tends towards a constant value.
If one assumes, as in the dynamical reduction models [6], that the collapse of the wavefunction is an abrupt and stochastic process in which a superposition abruptly reduces at a point in time to one of its states, and where the stochastic probabilities of these events is given, as described by equations (8), by the superposition's decay rate $\Gamma(t)$ and by Born's rule, we expects for each measurement a different course of $\dot{N}(t)$ as shown in the middle part of figure 2. At the beginning, all traces initially follow the curve $\dot{N}_{sup}(t)$ and then jump abruptly to either $\dot{N}_{dc}(t)$ or $\dot{N}_0$, as shown in the figure. The decay rate of the superposition $\Gamma(t)$ can be determined from these individual traces by averaging over all traces $\dot{N}_{\to dc}(t)$ jumping after some time to $\dot{N}_{dc}(t)$. The average of these traces $\langle \dot{N}_{\to dc} \rangle(t)$ can be calculated as follows. The average over all traces $\langle \dot{N} \rangle(t)$ is given by:

$$\langle \dot{N} \rangle(t) = \; P_{sup}(t)\, \dot{N}_{sup}(t) + P_{mov}(t)\, \dot{N}_{dc}(t) + P_{\neg mov}(t)\, \dot{N}_0(t)\; . \tag{10}$$

When averaging over all the traces that jump to $\dot{N}_{dc}(t)$ after some time, the last term in this expression must be skipped and the term $P_{sup}(t)\,\dot{N}_{sup}(t)$ must be weighted with $|c|^2_{mov}$, which takes into account that only the part $|c|^2_{mov}$ of the traces belonging to this term will later jump to $\dot{N}_{dc}(t)$ according to Born's rule. Since the ensemble over which the mean is taken is now by the factor $|c|^2_{mov}$, smaller, we must divide the result by $|c|^2_{mov}$, which leads to:

$$\langle \dot{N}_{\to dc} \rangle(t) = \frac{1}{|c|^2_{mov}} \left( |c|^2_{mov}\, P_{sup}(t) \dot{N}_{sup}(t) + P_{mov}(t) \dot{N}_{dc}(t) \right) . \tag{11}$$

With $\dot{N}_{sup}(t) = |c|^2_{mov} \dot{N}_{dc}(t) + |c|^2_{\neg\, mov} \dot{N}_0$, $|c|^2_{mov} + |c|^2_{\neg\, mov} = 1$, and solving equation (8)[1], this equation can be transformed to:

$$\langle \dot{N}_{\to dc} \rangle(t) = \dot{N}_{dc}(t) - |c|^2_{\neg\, mov} (\dot{N}_{dc}(t) - \dot{N}_0)\, e^{-\int_0^t dt \Gamma(t)} . \tag{12}$$

The right-hand side of figure 2 shows an example of a course of $\langle \dot{N}_{\to dc} \rangle(t)$. With result (12), the decay rate of the superposition $\Gamma(t)$ can be determined from the measured courses $\langle \dot{N}_{\to dc} \rangle(t)$, $\dot{N}_{dc}(t)$ and $\dot{N}_0$, which allows us to test collapse models.

## 4. Experimental condition for measuring collapse

In this section we discuss the experimental condition required for a decoherence-free measurement of wavefunction collapse.

Our experiment is based on the idea that the quantum superposition of the mirror changes the probability of detecting a photon in the Michelson interferometer, but conversely that this photon detection does not affect the quantum superposition of the mirror. This changes when the single-photon detection in the Michelson interferometer is replaced by a classical intensity measurement of a continuous light beam in the interferometer. In this case, if the mirror evolves into a superposition of the states *'mirror moved'* and *'mirror not moved'*, the detector of the Michelson interferometer also becomes part of this two-state superposition, since it measures different intensities in these states. As soon as the classical detector is able to reduce the two-state superposition, the measured intensity curve $I(t)$ will follow either the dark-count curve $I_{dc}(t)$ or the zero curve $I_0$, making a measurement of the quantum superposition of the mirror impossible. The crucial condition for measuring the superposition of the mirror is that the measurement in the Michelson interferometer is not a classical one, but directly or indirectly a measurement of reduction probabilities of superposed interferometer states. In our experiment, the detection of each photon is (in an indirect way) a measurement of the reduction probability that the superposition of the interferometer, consisting of the *'photon detected'* and *'photon not detected'* state, reduces to the '*photon detected'* state.
However, the experiment does not necessarily require the detection of individual photons. Even when, for example, the intensity of the light beam in the Michelson interferometer is recorded using a digital oscilloscope, the measured outcome can arise from the collapse of a superposed state involving the oscilloscope and the detector. We will return to this option of the experiment in section 5.

[1] From equations (8) follows: $P_{sup}(t) = e^{-\int_0^t dt\Gamma(t)}$ and $P_{mov} = |c|^2_{mov}(1 - e^{-\int_0^t dt\Gamma(t)})$.

# 5. Technical implementation of the experiment

In this section we discuss how the proposed experiment can be realised using currently available technologies for photon generation and detection.

## 5.1. Generation of the quantum superposition of the mirror

To generate the quantum superposition of the mirror via a single-photon measurement in figure 1, wavelengths in the visible range (e.g., $633nm$) can be used, where the photon can be detected using low-cost Si-SPAD avalanche photodiodes, which also offer high quantum efficiencies. The single photon can be generated using a laser attenuated to the single-photon level. However, it is better to use techniques such as spontaneous parametric down-conversion (SPDC) or quantum dots, where the generation of a single photon is guaranteed. The use of quantum dots offers the advantage of significantly higher photon generation efficiencies (50-80%) than the other methods (1-10% for attenuated lasers and <1% for SPDC) but requires more experimental effort due to the need for low temperatures.

To measure the $\langle \dot{N}_{\rightarrow dc} \rangle(t)$ -curve, it is important to know the start time of the experiment. For attenuated lasers and quantum dots, this is given by the time of the laser pulse, and for SPDC by the detection of the signal photon.

Whether a measured curve $\dot{N}(t)$ hast to be taken into account for the calculation of the averaged curve $\langle \dot{N}_{\rightarrow dc} \rangle(t)$, which is the case when the mirror moves and $\dot{N}(t)$ approaches to $\dot{N}_{dc}(t)$, can be determined most easily by measuring the voltage at the piezo actuator, which indicates whether the piezo actuator has been charged by the avalanche current from the photodiode or not. However, to prevent this voltage measurement from disturbing the superposition of the mirror, it must be carried out at enough time after the mirror superimposition has been reduced.

## 5.2. Photon detection in the Michelson interferometer

Wavelengths in the visible range can also be used for the photon detection in the Michelson interferometer, which are preferably detected with Si-SPADs having higher quantum efficiencies (up to 70%) than photomultipliers tubes (10–30%). In order to fulfil conditions (4) and (5), it is important to control the lifetime of the photon detection $T_{ph-det}$ and ensure that it is as short as possible. According to the Diósi-Penrose criterion (cf. section 6), this lifetime can be shortened when the avalanche current of the photodiode drives a piezo actuator that moves a body with a large mass. The calculations in reference [15], show that, with this technique, lifetimes of about $0.1\mu s$ can be achieved. This reduction time is short enough to satisfy the condition $T_{ph-det} \ll T_{mirr}$ (4) for the example calculations in section 6.4, where the mirror collapses at approximately $0.5\mu s$ for $R = 1k\Omega$ and around $4\mu s$ for $R = 25k\Omega$ (see figure 4). From the condition $T_{ph-det} \ll T_{in}$ (5) it follows that the laser emission rate $\dot{N}_{in}$ should not exceed $10MHz$ ($T_{in} = 1/\dot{N}_{in}$). Photon rates of this magnitude can be measured with Si-SPADs.

The photons in the Michelson interferometer can be generated by the same techniques used to generate the quantum superposition of the mirror, such as attenuated lasers, spontaneous parametric down-conversion (SPDC) or quantum dots (see section 5.1). These techniques allow photon generation rates of $10MHz$ and higher.

In the following, we discuss three options for how the decay of the quantum superposition of the mirror can be measured with the Michelson interferometer.

### Option 1: Measurement of the photon detection rate

This is the method already described in sections 2 and 3, where a continuous stream of photons is sent into the Michelson interferometer. With this method, photon rates $\dot{N}_{in}$ of up to $10MHz$ can be used, as discussed above. For the sample calculations in section 6.4, this would allow for approximately five photon detections per run, when the mirror collapses at $0.5\mu s$ for $R = 1k\Omega$ (see left part of figure 4), or for forty photon detections per run, when the mirror collapses at $4\mu s$ for $R = 25k\Omega$ (see right part of figure 4).

### Option 2: Detection of one photon per run

If there is a concern that the repeated measurement of photons in the Michelson interferometer in option 1 might affect the superposition of the mirror, one can minimise this interaction by detecting only one photon per run. This photon is then sent into the interferometer at a dedicated time $t$, which can be varied. The mean probability of detecting the photon for all runs at which the mirror superposition finally collapses to the *'mirror moved'* state $\langle p_{\rightarrow mov} \rangle$ is given by:

$$\langle p_{\rightarrow mov} \rangle(t) = p_{ph}(\Delta s(t)) - |c|^2_{\neg mov} \left(p_{ph}(\Delta s(t)) - p_{ph}(0)\right) e^{-\int_0^t dt \Gamma(t)} , \qquad (13)$$

which follows directly from equation (12) by dividing it by $\dot{N}_{in}$. For this option, photon generation with quantum dots is most suitable due to the significantly higher photon generation efficiency compared to attenuated lasers or SPDC (cf. section 5.1).

It is worth noting that for this option, unlike a classical measurement, the result of the photon detection does not allow any conclusions to be drawn about the state of the mirror, whether it is still in superposition or has already been reduced to one of the states.

### Option 3: Intensity measurement of a light beam

If, instead of single photons, a continuous light beam is sent into the Michelson interferometer and its intensity $I$ is measured with e.g. a solid-state detector and recorded on a digital oscilloscope, the result displayed on the oscilloscope can result from a reduction of a superposed state within the oscilloscope. In this case, we do not have a classical intensity measurement as discussed in section 4. The reduction of the state within the oscilloscope is then decoupled from the reduction of the mirror superposition, as in the case of single photon detection. Whether this assumption holds for a given oscilloscope can be seen from the measured results, when a difference between the $\langle I_{\rightarrow dc} \rangle(t)$- and the $I_{dc}(t)$-curve occurs.

Since we do not know how fast a superposed state within the oscilloscope decays, we don't know how to choose the sampling rate of the oscilloscope to prevent successive samples from being in superposition at the same time (cf. condition (5)). This problem can be avoided by measuring the intensity of the light beam at only one point in time $t$ for each run, as in option 2.

### 5.3. Measurement of the dark-count curve

For the evaluation of the experiment with equations (12) or (13), it is necessary to measure the dark count curve $\dot{N}_{dc}(t)$, or the dark count probability over time $p_{ph}(\Delta s(t))$. This can be done by waiting for dark count events, where the occurrence and the start time of the event can be determined by measuring the voltage at the piezo actuator. Alternatively, the avalanche breakdown in the photodiode can be triggered using a light pulse.

# 6. Expected outcomes for the gravity-based collapse models

In this section, we calculate the results predicted for our experiment by the gravity-based collapse models of Diósi and Penrose [5] to demonstrate the feasibility of the experiment for testing the established collapse models, which include, in addition to the gravity-based models, the dynamical reduction models such as Continuous Spontaneous Localization [6].
In section 6.1 we introduce the models of Diósi and Penrose, in section 6.2 we propose refinements to our experiment, in section 6.3 we present the formulas for calculating the decay rate of the mirror and the piezo actuator that moves the mirror, and in section 6.4 we calculate the expected results, referring to commercially available components for.

### 6.1. Diósi-Penrose criterion

The Diósi-Penrose criterion [7,8] is a rule of thumb resulting from the gravity-based models of Diósi [9,10] and Penrose [11,12]. The decay rate $\Gamma$ of a two-state superposition is described in these gravity-based models by a characteristic gravitational energy between the states $E_{DP}$:

$$\Gamma(t) = \frac{\gamma E_{DP}(t)}{\hbar}, \qquad (14)$$

where $\gamma$ is a dimensionless factor which is expected to be of the order of one [12] and which is chosen to be one in the calculations in section 6.4. The energy $E_{DP}$, which we refer to here as the *Diósi-Penrose energy*, depends on the difference of the mass density distributions $\rho_1(\boldsymbol{x})$ and $\rho_2(\boldsymbol{x})$ of the superposed states and is described by minus the gravitational self-energy resulting from this difference, which is given by [12,5,13,14]:

$$E_{DP} = \frac{G}{2}\int d^3\boldsymbol{x} d^3\boldsymbol{y} \frac{(\rho_2(\boldsymbol{x}) - \rho_1(\boldsymbol{x}))(\rho_2(\boldsymbol{y}) - \rho_1(\boldsymbol{y}))}{|\boldsymbol{x} - \boldsymbol{y}|}\ , \qquad (15)$$

where $G$ is the gravitational constant. The Diósi-Penrose energy is always positive, which follows directly from Penrose's approach [15].

The models of Diósi and Penrose are not completely identical in their predictions, since in Diósi's model for the calculation of the mass density distributions of the states $\rho_i(\boldsymbol{x})$ a spatial averaging must be introduced with a smearing radius in the order of $10^{-5} cm$ [10,16]. This averaging eliminates the microscopic mass distribution of solids in which the mass is localised at the locations of the atomic nuclei. In our experiment, however, this averaging is only relevant when the displacement of the mirror $\Delta s$ becomes smaller than about 10 lattice constants [15]. Diósi's model will hereafter be referred to as the *Diósi-Penrose model with smeared mass-density operator* and Penrose's model as the *parameter-free Diósi-Penrose model*.

### 6.2. Refinements of the experiment

When calculating the decay rate of the superposition of our experiment $\Gamma$ with the Diósi-Penrose energy $E_{DP}$ (14), it is important to consider that not only the mirror and the piezo have different mass distributions $\rho(\boldsymbol{x})$ in the two superposed states, which must be taken into account when calculating the Diósi-Penrose energy with equation (15), but also the power supply in figure 1. Since a current has only flowed out of the power supply in the *'photon detected'* state, slightly different mass density distributions $\rho(\boldsymbol{x})$ are expected for the *'photon detected'* and the *'photon not detected'* state in the power supply, which must be taken into account when calculating the Diósi-Penrose energy. The same applies to the photodiode and the wires connecting the components in figure 1. In reference [15], formulas are derived to estimate the Diósi-Penrose energy of the photodiode and the wires when a current has flowed through them in one of the states. The calculation the Diósi-Penrose energy for the power supply is difficult because it consists of many components and is connected to the power grid. This problem can be avoided by replacing the power supply with a simple plate capacitor that is charged to the required voltage by a conventional power supply before the measurement, and by disconnecting the power supply from the plate capacitor during the measurement by opening the two switches in figure 3. The estimates of the Diósi-Penrose energies of the plate capacitor, the photodiode and the wires in reference [15] show that they are much smaller than the Diósi-Penrose energies of the mirror and the piezo, which is why they are neglected in the example calculations in section 6.4.

For our experiment, it is also recommended to be able to control the charging time of the piezo actuator by connecting a resistor $R$ in series, as shown in figure 3. The example calculations in section 6.4 will show that by delaying the displacement of the mirror, it is possible to examine areas in which differences between the parameter-free and the Diósi-Penrose model with smeared mass-density operator are expected.

Another interesting extension of the proposed experiments, not shown in figure 3, is to use the avalanche current from the photodiode to control another piezo actuator that simply displaces a mass whose mass density and size can be varied. The Diósi-Penrose energy of this piezo actuator then additionally contributes to the decay of the superposition. This extension is of particular interest for testing the gravity-based collapse models, where the decay rate increases with the square of the mass density of the displaced mass [15].

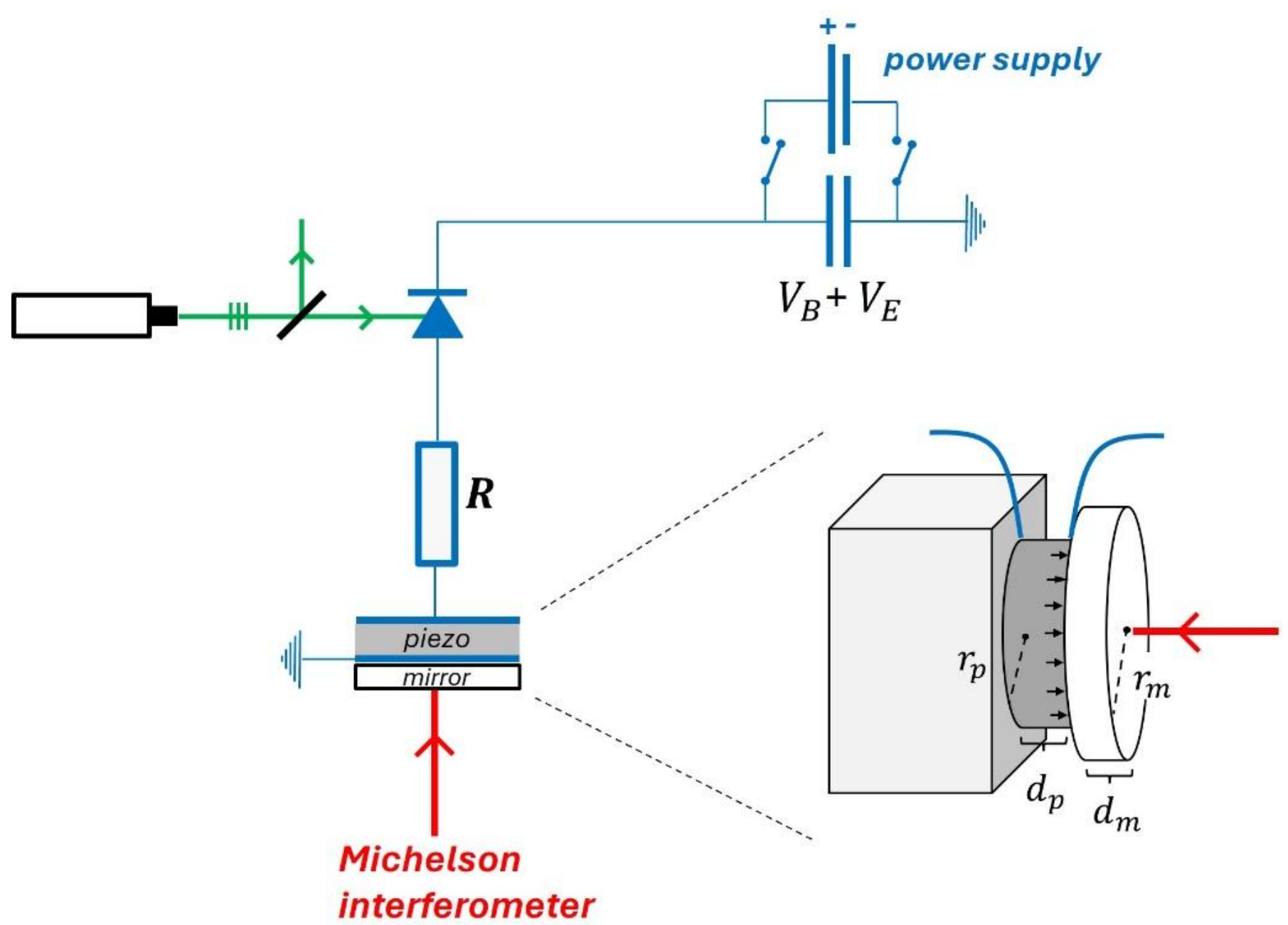

**Figure 3.** Refinements of the setup in figure 1.

1) Replacement of the conventional power supply by a plate capacitor, which is charged shortly before the photons arrival by a conventional power supply and disconnected from this power supply during the measurement.
2) Resistor $R$ in front of the piezo actuator delaying the displacement of the mirror.

### 6.3. Diósi-Penrose energies of the components

In reference [15], formulas were developed for calculating the Diósi-Penrose energy of superposed solids that are either slightly displaced relative to each other or have slightly different expansions in the states. The formulas were also derived for a displaced mirror and the piezo actuator in figure 3, which displaces the mass to the right when a voltage is applied. When the piezo actuator has displaced the mass by $\Delta s$ in one state and not in the other, the Diósi-Penrose energy for the superposed mirror and the piezo actuator is given by:

$$E_{DP} \approx 2\pi G \left( \frac{1}{3}\rho_p^2 d_p \pi r_p^2 \left(1 + 0.64\frac{d_p}{r_p}\right) + \rho_m^2 d_m \pi r_m^2 \left(1 + 0.64\frac{d_m}{r_m}\right) \right) \Delta s^2 \; . \qquad (16)$$

Here $\rho_p$, and $\rho_m$ are the mass densities of the piezo and of the mirror, respectively, while the dimensions $d_p$, $d_m$, $r_p$ and $r_m$ are defined in figure 3.

Equation (16) assumes a homogeneous mass distribution within the solids, neglecting the discrete localization of mass at the atomic nuclei, and therefore applies exactly to the Diósi-Penrose model with smeared mass-density operator (cf. section 0). When the microscopic mass distribution of the solid is considered within the parameter-free Diósi-Penrose model, the resulting Diósi-Penrose energy $E_{DP}$ increases for displacements $\Delta s$ smaller than approximately 10 lattice constants $g$ of the solid [15], where $g$ typically is about two Ångströms. This increase becomes dramatic when the displacement $\Delta s$ is smaller than the spatial variation $\sigma$ of the atomic nuclei, which is determined by the excited acoustic phonons in the solid and is typically on the order of a tenth of an Ångström at room temperature [15]. In this regime ($\Delta s < \sigma$), the Diósi-Penrose energy in the parameter-free model exceeds that of the smeared mass-density operator model by a material-dependent factor $\xi$, which is at room temperature typically around two orders of magnitude and increases further at lower temperatures [15].

### 6.4. Example calculations

For the following example calculations, we first need to calculate how the mirror moves over time $\Delta s(t)$ when a photon triggers an avalanche current in the photodiode in figure 3. To do this, we need to know the voltage curve at the piezo actuator $V(t)$, which depends on the excess bias voltage $V_E$ applied to the photodiode[2], the internal resistance of the photodiode $R_{di}$, the capacitance of the piezo actuator $C_p$[3] and the resistance $R$ in figure 3 as follows [17,15]:

$$V(t) = V_E\left(1 - e^{-\frac{t}{\tau_p}}\right) \qquad \text{with: } \tau_p = (R + R_{di})C_p \; . \qquad (17)$$

The displacement of the mirror $\Delta s(t)$ depends on the voltage at the piezo actuator as follows [15]:

[2] The voltage applied to the photodiode is $V_B + V_E$ as shown in figure 3, where $V_B$ is the breakdown voltage of the photodiode.

[3] The capacitance of the piezo actuator is: $C_p = \varepsilon_0 \varepsilon_r \pi r_p^2 / d_p$, where $\varepsilon_r$ is the relative permittivity of the piezo and $r_p$ and $d_p$ are defined in figure 3.

$$\Delta s(t) = d_{33}\, V(t) \quad , \tag{18}$$

where $d_{33}$ is a component of the matrix for the converse piezoelectric effect, which describes the extension of the piezo in the z-direction when an electric field $E$ is applied in the same direction ($\Delta d/d = d_{33}E$).

With the above formulas, formulas (14) and (16) for the decay rate of the mirror and piezo actuator, and with result (12), the expected course of the $\langle \dot{N}_{\to dc}\rangle(t)$ -curve and the corresponding dark-count curve $\dot{N}_{dc}(t)$ can be calculated for specific mirrors, piezo actuators, photodiodes, and for given parameters such as the excess bias voltage $V_E$ and the resistance $R$ in figure 3. Figure 4 shows the results of such calculations, which relate to commercially available components and common parameters specified in the lower part of the figure. The upper part shows the $\langle \dot{N}_{\rightarrow dc} \rangle(t)$ –curve and the corresponding dark-count curve $\dot{N}_{dc}(t)$, where the percentage deviation from the zero curve $\dot{N}_0$ is displayed $(\langle \dot{N}_{\rightarrow dc} \rangle - \dot{N}_0)/\dot{N}_0)$. The middle part of the figure shows the corresponding displacement curves of the mirror $\Delta s(t)$. The calculations of the $\langle \dot{N}_{\rightarrow dc} \rangle(t)$-curves were performed for both, the parameter-free Diósi-Penrose model and the model with smeared mass-density operator. The additional contribution to the Diósi-Penrose energy, which arises when the microscopic mass distribution of the solid is considered within the parameter-free Diósi-Penrose model (cf. section 6.3) and which must be added to result (16), were calculated with the formulas in reference [15] for room temperature.

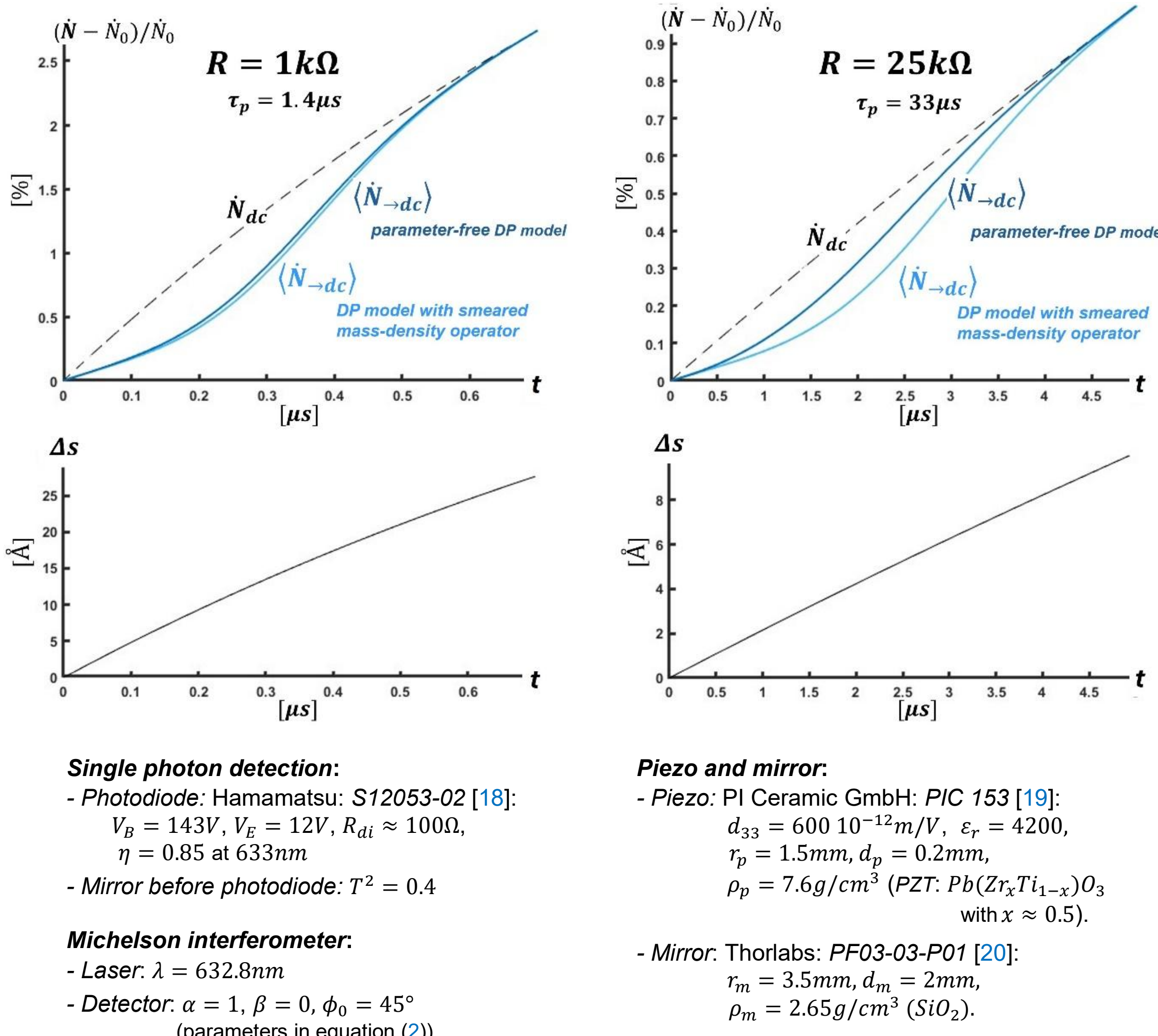

**Figure 4.** Example calculations for the components and parameters specified at the bottom. The upper diagrams show the difference between the $\langle \dot{N}_{\to dc}\rangle$-curves and the dark-count curves $\dot{N}_{dc}$, and the lower ones the course of the corresponding mirror displacement $\Delta s(t)$. In the calculations on the right, the resistance in figure 3 was increased to $R = 25k\Omega$, which leads to smaller mirror displacement $\Delta s(t)$, making differences between the parameter-free Diósi-Penrose model and the model with smeared mass density operator apparent.

The calculation on the right in figure 4 has been performed for a larger resistance $R = 25k\Omega$ (cf. figure 3) to slow down the mirror motion $\Delta s(t)$, which then remains far below ten lattice constants ($\approx 20\text{Å}$), for which the parameter-free Diósi-Penrose model predicts larger Diósi-Penrose energies and correspondingly larger decay rates (cf. section 6.3). Here the difference between the parameter-free Diósi-Penrose model and the model with smeared mass density operator becomes apparent.

The mirror and piezo considered in our example calculation in figure 4, which are both in a quantum superposition in the experiment, have together a weight of $0.2\ g$ and consist of approximately $6 \cdot 10^{21}$ atoms.

# 7. Discussion

In this paper, it was shown that with interferometers in quantum superpositions the decay of quantum superpositions involving $6 \cdot 10^{21}$ atoms can be measured, which is sufficient for testing the established collapse models. Further, it was shown that the proposed experimental agenda can realised with currently available technology for photon generation and detection.

The key advantage of using interferometers in quantum superpositions to measure collapse - compared to the conventional interferometric method - is that it provides direct feedback on whether the collapse of the mirror superposition is detectable with the chosen equipment, namely when differences between the $\langle \dot{N}_{\rightarrow dc} \rangle(t)$- and the $\dot{N}_{dc}(t)$-curve occur. In the interferometric method, there is always an inherent uncertainty as to whether the disappearance of quantum interference between superposed states is due to a genuine collapse or merely the result of decoherence.

Another advantage of the proposed method is that, due to the absence of decoherence, measurements can be performed at room temperature, which significantly reduces the experimental effort and complexity.

Initiating the proposed experimental agenda should be feasible for quantum optics groups with relatively little effort. To generate the quantum superposition of the mirror with the single-photon measurement in figure 1, the required photon can initially be generated using attenuated lasers. For the measurement in the Michelson interferometer, one can begin with option 3 in section 5.2, where, instead of single photons, the intensity of a continuous light beam is measured, which can be generated with e.g. a conventional HeNe laser. This initial setup can be used to test for which of the available detectors and recording devices differences between the $\langle I_{\rightarrow dc} \rangle(t)$- and the $I_{dc}(t)$-curve occur, which, if positive, would allow for a first measurement of wavefunction collapse. For a more precise analysis, the measurements should be done with single photons, as described in options 1 and 2 of section 5.2.

## Acknowledgement

I am very grateful to James Tagg for the many discussions about his experimental proposal.

# Appendix: Experiment of Tagg

In this appendix, we present Tagg's experimental proposal [4] and show how it can be evaluated with the results derived in this work, if we assume that single photons are detected in the interferometer.
Figure 5 illustrates the setup, where mirrors in both arms of a Mach-Zehnder interferometer are controlled by piezo actuators. The single-photon measurement in figure 5, which splits the photon from the single-photon laser into two beams, ensures that only one mirror is moved at a time.

The probability of measuring a photon with the detector of the Mach-Zehnder interferometer depends on the mirror displacements $\Delta s_+(t)$ and $\Delta s_-(t)$ (see figure 5) as follows:

$$p_{ph}(\Delta s_+, \Delta s_-) = \alpha\, sin^2\left(4\pi\frac{\Delta s_+ - \Delta s_-}{\lambda} + \phi_0\right) + \beta \quad . \tag{19}$$

The factor $4\pi$ instead of $2\pi$ in equation (2) is related to the fact that the photon is reflected twice at the mirrors because of the corner cube reflectors that are needed to shift the beam to the side to close the square. Since dark count events can occur in both photodiodes, we get two dark-count curves, $\dot{N}_{dc+}(t)$ and $\dot{N}_{dc-}(t)$, in which the photon rate either swings upwards or downwards, as shown in figure 5. These dark count curves and the zero curve $\dot{N}_0$ can be calculated as follows:

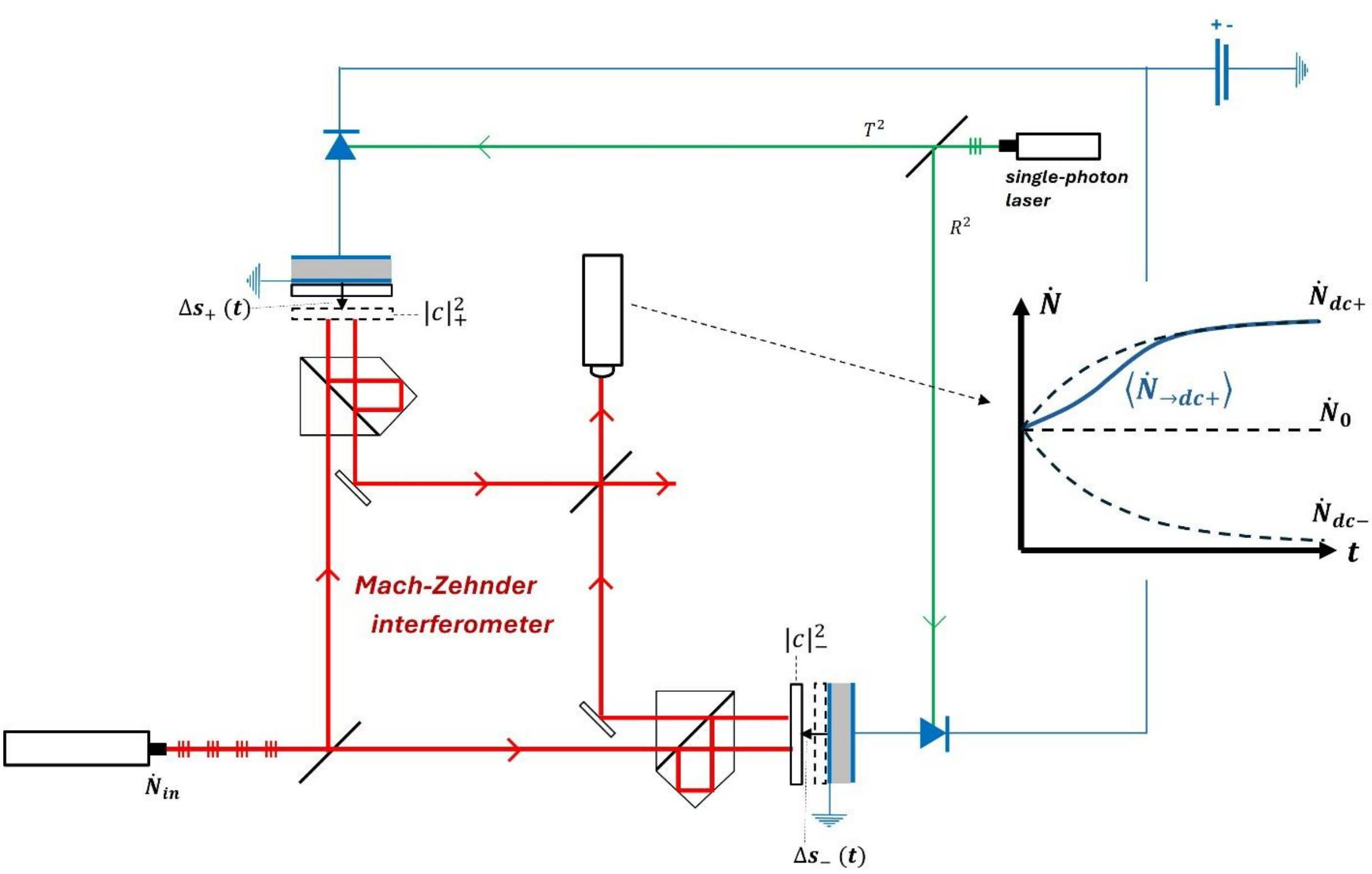

**Figure 5.** Tagg's experiment, in which the detection of a split single photon causes either the upper or right mirror of the Mach-Zehnder interferometer to be displaced, resulting in an increase or decrease in the photon detection rate $\dot{N}$, as shown on the right.

$$
\begin{aligned}
\dot{N}_{dc+}(t) &= \dot{N}_{in}\, p_{ph}(\Delta s_{+}(t), 0) \\
\dot{N}_{0}(t) &= \dot{N}_{in}\, p_{ph}(0,0) \qquad . \\
\dot{N}_{dc-}(t) &= \dot{N}_{in}\, p_{ph}(0, \Delta s_{-}(t))
\end{aligned}
\tag{20}
$$

Since the split photon is detected by photodiodes with finite quantum efficiencies $\eta_{+}$ and $\eta_{-}$, four possible states must be considered: the photon either passes through or is reflected by the beam splitter and is subsequently either detected or not detected by the corresponding photodiode. The intensities of the states in which the photon was detected by the upper or right photodiode, $|c|_{+}^{2}$ and $|c|_{-}^{2}$, or was not detected at all, $|c|_{0}^{2}$, are given by:

$$
\begin{aligned}
|c|_{+}^{2} &= T^{2}\eta_{+} \\
|c|_{-}^{2} &= R^{2}\eta_{-} \qquad , \\
|c|_{0}^{2} &= 1 - T^{2}\eta_{+} - R^{2}\eta_{-}
\end{aligned}
\tag{21}
$$

where $T^{2}$ and $R^{2}$ are the transmission and reflection coefficients of the beam splitter (see figure 5). When calculating the average over all traces $\langle \dot{N}_{\rightarrow dc+} \rangle(t)$ jumping after some time to the dark-count curve $\dot{N}_{dc+}(t)$ (see figure 5), it must be taken into account that the upper and the right mirrors have their own decay rates $\Gamma_{+}(t)$ and $\Gamma_{-}(t)$, which reduce the superposition independently of each other. Consequently, the initial superposition can decay via different paths to the state in which the photon was detected by the upper detector.
Considering that the power supply at the top right of figure 5 also has a decay rate $\Gamma_{ps}(t)$, which describes the decay of the superposition where an avalanche current is either drawn by one of the photodiodes or not, we arrive—after a lengthy calculation—at the following result for $\langle \dot{N}_{\rightarrow dc+} \rangle(t)$:

$$
\begin{aligned}
\langle \dot{N}_{\rightarrow dc+} \rangle(t) = {} & \dot{N}_{dc+}(t) \\
& - |c|_{-}^{2}\left(\dot{N}_{dc+}(t) - \dot{N}_{dc-}(t)\right)\Big[ e^{-\int_0^t dt(\Gamma_{+}(t)+\Gamma_{-}(t)+\Gamma_{ps}(t))} + \\
& \qquad \frac{1}{|c|_{+}^{2}+|c|_{-}^{2}}\left(1 - e^{-\int_0^t dt\Gamma_{ps}(t)}\right) e^{-\int_0^t dt(\Gamma_{+}(t)+\Gamma_{-}(t))} \Big] \\
& - |c|_{0}^{2}\left(\dot{N}_{dc+}(t) - \dot{N}_{0}\right)\Big[ e^{-\int_0^t dt(\Gamma_{+}(t)+\Gamma_{-}(t)+\Gamma_{ps}(t))} + \\
& \qquad \frac{1}{|c|_{0}^{2}+|c|_{+}^{2}}\left(1 - e^{-\int_0^t dt\Gamma_{-}(t)}\right) e^{-\int_0^t dt(\Gamma_{+}(t)+\Gamma_{ps}(t))} \Big].
\end{aligned}
\tag{22}
$$

Tagg proposed a variant for his experiment in which the displacement of one mirror is inverted ($\Delta s_{-}(t) = -\Delta s_{+}(t)$) [4]. In this case we get $\dot{N}_{dc+}(t) = \dot{N}_{dc-}(t)$, and only the term proportional to $|c|_{0}^{2}$ in equation (22) remains.

## References

[1] ***Fein, Y.Y., Geyer, P., Zwick, P. et al.***, Quantum superposition of molecules beyond 25 kDa, *Nat. Phys.* 15, *1242–1245 (2019),* *https://doi.org/10.1038/s41567-019-0663-9*

[2] ***Friedman, J., Patel, V., Chen, W. et al.***, Quantum superposition of distinct macroscopic states, *Nature* 406, *43–46 (2000),* *https://doi.org/10.1038/35017505*

[3] ***B. Schrinski, Y. Yang, U. von Lüpke,2 M. Bild,2 Y. Chu, K. Hornberger, S. Nimmrichter, M. Fadel***, Macroscopic Quantum Test with Bulk Acoustic Wave Resonators, *Phys. Rev. Lett.* 130, *133604 (2023), arXiv:2209.06635*

[4] ***J. Tagg, W. Reid, D. Carlin***, Schrödinger's Cheshire Cat: A tabletop experiment to measure the Diósi-Penrose collapse time and demonstrate Objective Reduction (OR), *(2024), arXiv:2402.02618*

[5] ***S. Donadi, A. Bassi***, Seven non-standard models coupling quantum matter and gravity, *AVS Quantum Sci., 4, 025601 (2022), arXiv:2202.13542*

[6] ***A. Bassi, G. C. Ghirardi***, Dynamical reduction models, *Phys. Rept., 379, 257 (2003)*, *arXiv:quant-ph/0302164*

[7] ***S. L. Adler***, Comments on proposed gravitational modifications of Schrodinger dynamics and their experimental implications, *J. Phys. A, 40, 755-764 (2007), arXiv:quant-ph/0610255*

[8] ***S. Gao***, On Diósi-Penrose criterion of gravity-induced quantum collapse, *Int. J. Theor.Phys., 49, 849-853, 2010, arXiv:1001.4857*

[9] ***L. Diósi***, Models for universal reduction of macroscopic quantum fluctuations, *Phys. Rev. A, 40, 1165-1174 (1989)*

[10] ***G. C. Ghirardi, R. Grassi, A. Rimini***, Continuous-spontaneous-reduction model involving gravity, *Phys. Rev. A, 42, 1057-1064 (1990)*

[11] ***R. Penrose***, On gravity's role in quantum state reduction, *Gen. Rel. Grav., 28, 581-600 (1996)*

[12] ***R. Howl, R. Penrose, I. Fuentes***, Exploring the unification of quantum theory and general relativity with a Bose-Einstein condensate, *New J. Phys., 21, 043047 (2019), arXiv:1812.04630*

[13] ***A. Bassi, A. Großardt, H. Ulbricht***, Gravitational decoherence, *Class. Quantum Grav., 34, 193002 (2017), arXiv:1706.05677*

[14] ***L. Diósi***, On the conjectured gravity-related collapse rate $E_\Delta/\hbar$ of massive quantum superpositions, *AVS Quantum Sci. 4, 015605-(4) (2022)*, *arXiv:2111.04604*

[15] ***G. Quandt-Wiese***, Diósi-Penrose criterion for solids and electrical components in quantum superpositions and application to a single-photon detector, *(2017), arXiv:1701.00353*

[16] ***A. Vinante, H. Ulbricht*****,** Gravity-related collapse of the wave function and spontaneous heating: revisiting the experimental bounds, *AVS Quantum Sci. 3, 045602 (2021)*, *arXiv:2109.14980*

[17] ***S. Cova, S. Ghioni, A. Lacaita, C. Samori, F. Zappa***, Avalanche photodiodes and quenching circuits for single-photon detection, *Applied Optics, 35(12), 1956-1976 (1996)*

[18] ***Hamamatsu***, Datasheet for "S12053-02", *www.hamamatsu.com*

[19] ***PI Ceramic GmbH***, Catalog on "Piezoelectric Ceramic Products", *http://www.piceramic.com*

[20] ***Thorlabs***, Datasheet for "PF03-03-P01", *www.thorlabs.com*

[21] ***M. Wayne, J. Bienfang, A. Migdall***, Low-noise photon counting above 100 million counts per second with a high-efficiency reach-through single-photon avalanche diode system, *Appl. Phys. Lett. 118, 134002 (2021)*, *https://doi.org/10.1063/5.0041984*

[22] ***A. Wright*****,** The photomultiplier handbook**,** *Oxford University Press (2017)*